\documentclass{aastex631}

\begin{document}

\title{Sun-as-a-star Analysis of the X1.6 Flare on 2023 August 5: Dynamics of Post-flare Loops in Spatially Integrated Observational Data}

\author{Takato Otsu}
\affiliation{Astronomical Observatory, Kyoto University, Sakyo, Kyoto, Japan}

\author{Ayumi Asai}
\affiliation{Astronomical Observatory, Kyoto University, Sakyo, Kyoto, Japan}

\author{Kai Ikuta}
\affiliation{Department of Multidisciplinary Sciences, The University of Tokyo, 3-8-1 Komaba, Meguro, Tokyo 153-8902, Japan}

\author{Kazunari Shibata}
\affiliation{Kwasan Observatory, Kyoto University, Yamashina, Kyoto 607-8471, Japan}
\affiliation{School of Science and Engineering, Doshisha University, Kyotanabe, Kyoto 610-0321, Japan}

\begin{abstract}
Post-flare loops are loop-like plasmas observed during the decay phase of solar flares, and they are expected to exist for stellar flares. However, it is unclear how post-flare loops are observed in stellar flares' cases. To clarify behaviors of post-flare loops in spatially integrated data, we performed the Sun-as-a-star analysis of the X1.6 flare that occurred on 2023 August 5, using GOES X-ray flux ($\sim10^7$ K), extreme ultraviolet (EUV) images taken by Atmospheric Imaging Assembly onboard the Solar Dynamic Observatory ($\ge10^{4.9}$ K) and H$\alpha$ data taken by Solar Dynamics Doppler Imager on board the Solar Magnetic Activity Research Telescope at Hida Observatory, Kyoto University ($\sim10^4$ K). As a result, this flare showed signatures corresponding to the important dynamics of the post-flare loops even in the spatially integrated data: (1) The H$\alpha$ light curve showed two distinct peaks corresponding to the flare ribbons and the post-flare loops. The plasma cooling in the post-flare loops generated different peak times in soft X-rays, EUV, and H$\alpha$ light curves. (2) Downflows were confirmed as simultaneous redshifted/blueshifted absorptions in the H$\alpha$ spectra. (3) The apparent rise of post-flare loops was recognized as a slowing of the decay for the H$\alpha$ light curve. These results are keys to investigating stellar post-flare loops with spatially integrated data. We also discuss the dependence of our results on flare locations and their possible applications to stellar observations.
\end{abstract}


\section{Introduction} \label{sec:intro}
Solar flares are sudden energy release in the solar atmosphere, and their dynamics are explained by the standard flare model \citep{ShibataMagara2011}. The energy released in the solar corona is transported to the chromosphere along magnetic loops. 
As a result of this, chromospheric evaporation occurs and loops are filled with hot plasma ($\sim10^7$ K). These hot loops are observed in soft X-rays. Then, the plasma cool to a lower temperature and the loops become visible in extreme ultraviolet (EUV; $\sim10^5-10^6$ K) bands to chromospheric lines such as H$\alpha$ ($\sim10^4$ K).
Such loop-structure plasma observed in main and decay phases of solar flares is called 'post-flare' loops\footnote{In this Letter we use "post-flare loops" according to convention, although some papers have recommended avoiding the term because "post-flare" loops are in fact natural part of evolving flares \citep{Svestka2007SoPh..246..393S}.}\citep[e.g.,][]{Kamio2003SoPh}.
Along post-flare loops, downflows of condensed plasma are often observed at transition-region and chromospheric 
temperatures. These cooled plasmas are thought to be formed by runaway radiative cooling, like the formation of quiescent coronal rains which occur in coronal loops on active regions \citep[in some literature, downflows along post-flare loops are called 'flare-driven coronal rains'; e.g.,][]{Antolin2022FrASS...920116A,Sahin2024ApJ...970..106S}.

Like the solar case, stellar flares have been observed on various types of stars \citep[e.g.,][]{Kowalski2024LRSP...21....1K}.
Some studies have suggested that stellar flares are associated with eruptive phenomena \citep[e.g.,][]{VeronigETAL2021,NamekataETAL2022a,InoueETAL2023ApJ,Namekata2024ApJ...961...23N,Notsu2024ApJ...961..189N}.
The Sun-as-a-star analyses --in which solar data are spatially integrated to be compared directly with stellar data-- support these possible observations of such stellar phenomena \citep{NamekataETAL2022a, OtsuETAL2022}.
Sun-as-a-star analyses of solar activity are keys to improving understanding of stellar flares and have been actively performed \citep{Ma2024ApJ...966...45M, Pietrow2024A&A...682A..46P, Otsu2024ApJ...964...75O, Leitzinger2024MNRAS.532.1486L}.

Like the eruptive phenomena, post-flare loops are also critical components of the flare model, and they are expected to be observed for stellar flares \citep{Heinzel2018ApJ...859..143H,Wollmann2023A&A...669A.118W}.
In observations of M dwarf flares, white-light curves observed by the Transiting Exoplanet Survey Satellite (TESS) sometimes exhibit secondary peaks after the main impulsive peaks \citep{Howard2022ApJ...926..204H}.
To investigate the mechanism of the secondary peaks in TESS flares, \citet{Yang2023ApJ...959...54Y} performed one dimensional hydrodynamic simulations along the flare loop and found that plasma condensation can lead to a secondary peak in synthetic light curves computed with the method of \citet{Heinzel2018ApJ...859..143H}. This suggests stellar post-flare loops can explain the secondary peak in TESS flares.
As other possible signatures of stellar post-flare loops, \citet{HondaETAL2018} reported redshifted absorption in H$\alpha$ spectra during the decay phase of an M dwarf flare, which can be interpreted as downflows along post-flare loops.
Also, stellar post-flare loops are proposed as possible causes of redshifted excess emission or red asymmetry in stellar H$\alpha$ spectra \citep{Wu2022ApJ...928..180W,NamizakiETAL2023ApJ,Wollmann2023A&A...669A.118W,Notsu2024ApJ...961..189N}.
For the further detailed investigation of stellar post-flare loops,
direct comparisons with solar post-flare loop data are essential as well as an approach via numerical simulations.
However, how stellar post-flare loops are observed in spatially integrated data is still ambiguous. Clarifying how post-flare loops are observed in spatially integrated solar data is required for detections and investigations of stellar post-flare loops.

In this Letter, we present our results of a Sun-as-a-star analysis for the X1.6 flare observed on 2023 August 5.
This flare showed typical post-flare loops in EUV and H$\alpha$ images.
The observation is presented in Section \ref{sec:obs}.
The methods for Sun-as-a-star analyses are described in Section \ref{sec:analysis}.
In Section \ref{sec:results}, we report our results, and provide discussions and conclusions in Section \ref{sec:discussion}.

\section{Observations} \label{sec:obs}
\subsection{Instruments}
We used H$\alpha$ spectral images taken by Solar Dynamics Doppler Imager \citep[SDDI;][]{IchimotoETAL2017} attached to the Solar Magnetic Activity Research Telescope \citep[SMART;][]{UenoETAL2004} at Hida Observatory, Kyoto University.
The SDDI takes full-disk solar images at 73 wavelength points between H$\alpha\pm9.0$ {\AA} with a spectral resolution of 0.25 {\AA}, a time cadence of 12 s, and pixel size of 1$^{\prime\prime}$.23, respectively.
Additionally, we used EUV images taken by Atmospheric Imaging Assembly \citep[AIA;][]{LemenETAL2012AIA} onboard the Solar Dynamic Observatory \citep[SDO;][]{PesnellETAL2012}.
The AIA takes full-disk solar images in EUV channels with a time cadence of 12 s and  pixel size of 0$^{\prime\prime}$.6.

\subsection{Event overview}
The target event is the X1.6 flare with the GOES flare peak at 22:21 UT on 2023 August 5 (Figure \ref{full} (a)), which occurred in the NOAA 13386 (Figure \ref{full} (b)) .
Figure \ref{dev} shows the time development of the flare in AIA 94, 171, and 304 {\AA}, and SDDI H$\alpha$ images.
During the impulsive phase of the flare, two ribbons were identified (Figure \ref{dev}, $t=70$ minutes from 2023 August 5 21:00 UT).
During the decay phase of the flare, bright post-flare loops dominantly appeared in the images of EUV and even the H$\alpha$ line center (Figure \ref{dev}, $t=120$ minutes).
Additionally, the dark features can be confirmed in the images of H$\alpha\pm1.0$ {\AA} (Figure \ref{dev} (d) and (f), $t=120$ minutes), and they are located along the post-flare loop.
These features are supposed to correspond to downflows along the post-flare loops.
The loops are located near the solar limb and are tilted against the line-of-sight direction. As a result, the downflows could be in both blue and red wings.
Later on, the upper part of the loops goes outside the solar disk due to the consecutive formation of higher and higher loops, as the result of a gradual reconnection (Figure \ref{dev}, $t=210$ minutes).

We made the time-slice diagram to describe the detailed dynamics of post-flare loops.
Figure \ref{TH} (a-1)-(d-1) show the time-slice diagrams of AIA channels and H$\alpha$ line center, with the vertical axis along the artificial slit from the white circle to the cross mark in Figure \ref{dev}.
The horizontal dashed line in the time-slice diagram corresponds to the limb indicated by the white dashed lines in Figure \ref{dev}. 
The GOES soft X-ray light curve and its peak time are also over-plotted as the dash-dot line and vertical dotted line, respectively.
Figure \ref{TH} (a-2)-(d-2) are the zoomed-in diagram corresponding to the regions indicated by the dashed square in Figure \ref{TH} (a-1)-(d-1).
In each time-slice diagram, flare ribbons are confirmed as two stripes at around 10-40 arcsec from the start point of the slit.
Almost at the same time as the GOES peak, the post-flare loops appeared in AIA 94 {\AA} (Figure \ref{TH} (a-1)-(a-2), t$\approx8$1 minutes). Subsequently, the loops appeared in cooler channels, i.e., AIA 171, 304 {\AA} (Figure \ref{TH} (b-2)-(c-2), t$\approx90$ minutes). Slightly after AIA 171 and 304, the loops appeared in H$\alpha$ (Figure \ref{TH} (d-2), t$\approx91$ minutes).
Higher loops appear with the apparent rising velocity of 4.5 km s$^{-1}$, which is consistent with white-light observations of an X-class flare reported by \citet{Jejcic2018ApJ...867..134J}. Finally, the upper part of the loops goes outside the solar disk at around $t=150$ minutes. We note that this scenario is similar to that presented in \citet{Svestka1987SoPh..108..237S}.

\section{Analysis} \label{sec:analysis}
We performed spatial integrations to investigate how post-flare loops could be observed on distant stars.
\subsection{SDO/AIA EUV data Analysis}
We made the light curves for the AIA channels of 94 {\AA}, 171 {\AA}, and 304 {\AA}, which have peak response temperatures of $\sim10^{6.8}$ K, $\sim10^{5.9}$ K, and $\sim10^{4.9}$ K, respectively \citep[e.g.,][]{Peter2012A&A...537A.152P}.
First, we integrated these data over the field of view shown in Figure \ref{dev}.
To focus on the signatures from the target event, we selected the local regions including the target X1.6 flare as the integral regions.
This restriction of the integral region equals to assuming that dominant temporal changes occurred only inside the selected region (Figure \ref{dev}).
Second, we subtracted pre-event data (2023-08-05 21:00 UT) from each integrated data. Finally, we normalized the integrated data with each peak value and obtained the spatially integrated and pre-event-subtracted AIA data.
\subsection{SMART/SDDI H$\alpha$ Data Analysis}\label{subsec:SDDI}
\begin{figure}[htbp]
\centering
\includegraphics[width=18cm]
{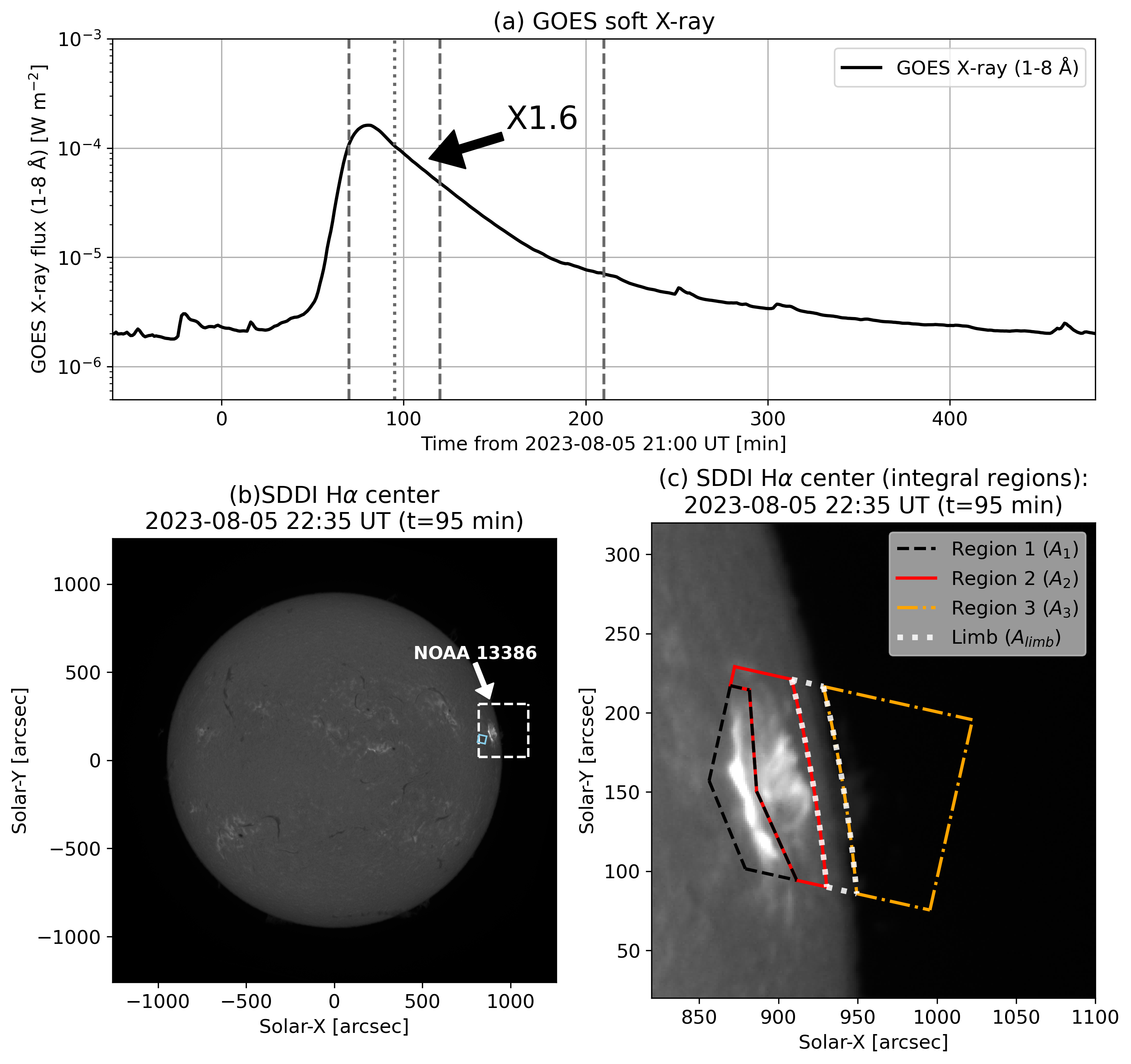}
\caption{(a) The GOES soft X-ray light curve between 20:00 UT on 2023 August 5 and 04:00 UT on 2023 August 6 is shown as black solid line. The black arrow indicates the target X1.6 flare. The vertical gray dotted line indicates the time of the panels (b) and (c) ($t=95$ minutes from 2023 August 5 21:00 UT). The vertical gray dashed lines indicate the times in Figure \ref{dev} ($t=70, 120$, and $210$ minutes).
(b) The H$\alpha$ line center solar full-disk image taken by SMART/SDDI at 22:35 UT on 2023 August 5. Solar north and west are at the top and right. The white arrow indicates the target active region NOAA 13386. The white dashed and skyblue boxes correspond to the field of view in Figure \ref{full} (c) and Figure \ref{dev}, and the quiet region for calibration, respectively.
(c) The integral and limb regions for H$\alpha$ analyses. The H$\alpha$ line center image taken by the SDDI at 22:35 UT on 2023 August 5 is shown with integral and limb regions. The black dashed, red solid, orange dash-dot, and white dotted regions correspond to the Region 1, 2, 3, and limb, respectively (see the text).}
\label{full}
\end{figure}

\begin{figure}[htbp]
\centering
\includegraphics[width=11cm]
{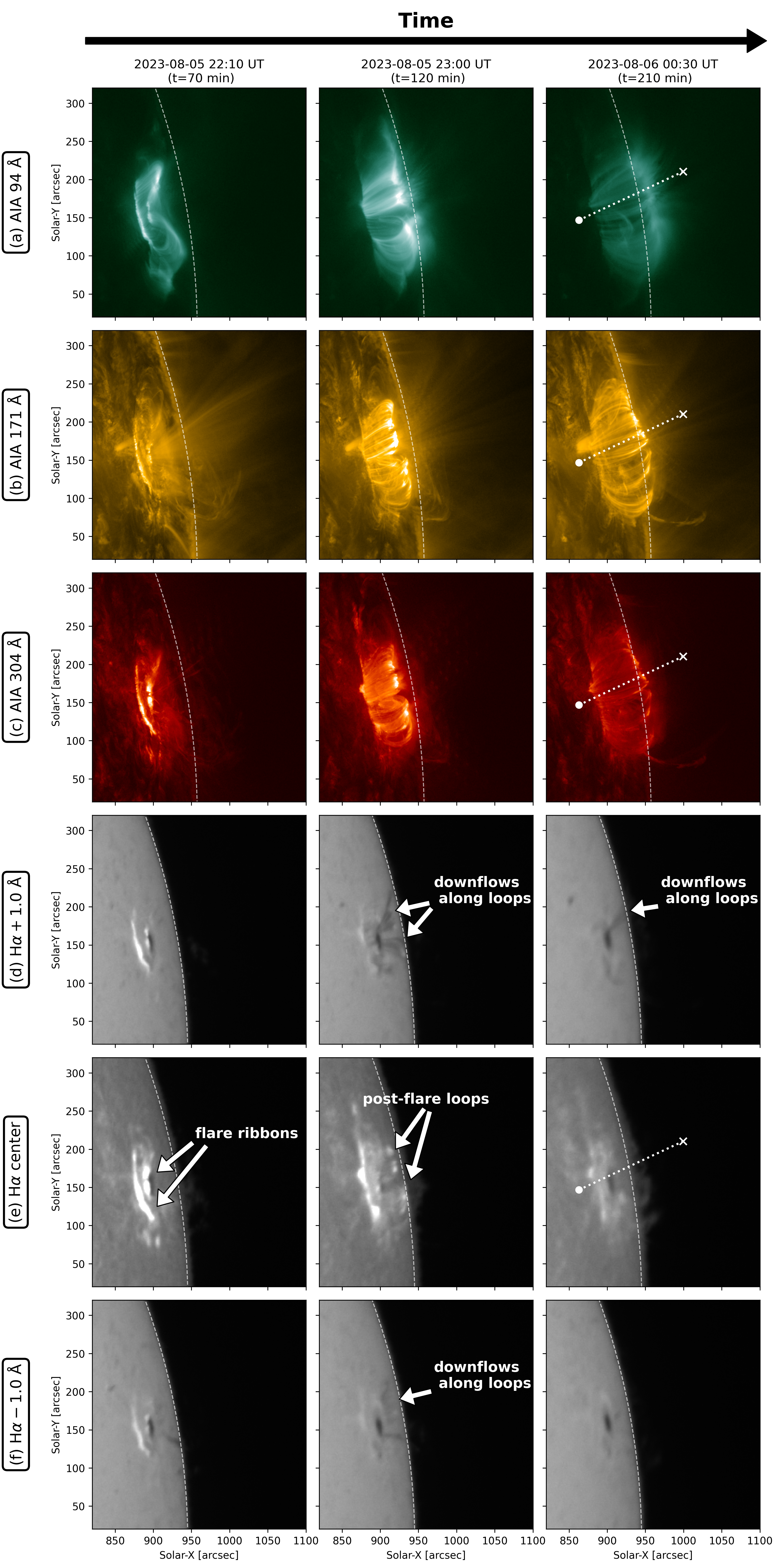}
\caption{Time development of the target event in EUV images and H$\alpha$ spectral images taken by SDO/AIA and SMART/SDDI, respectively. From left to right, images at 2023 August 5 22:10 UT ($t=70$ minutes from 2023 August 5 21:00 UT), 23:00 UT ($t=120$ minutes), and 2023 August 6 00:30 UT ($t=210$ minutes) are shown. In the top three rows, images of AIA 94 {\AA} (a), 171 {\AA} (b), and 304 {\AA} (c) are shown. In the bottom three rows, H$\alpha+1.0$ {\AA} (d), H$\alpha$ line center (e), and H$\alpha-1.0$ {\AA} are shown. The field of view of all the panels corresponds to the white dashed box in Figure \ref{full} (b).
The white dashed line in each panel indicates the limb.
Some notable points are indicated by the white arrows. The white dotted line connecting the white circle and cross is the artificial slit for the time-slice diagram in Figure \ref{TH}.}
\label{dev}
\end{figure}

\begin{figure}[htbp]
\centering
\includegraphics[width=18cm]
{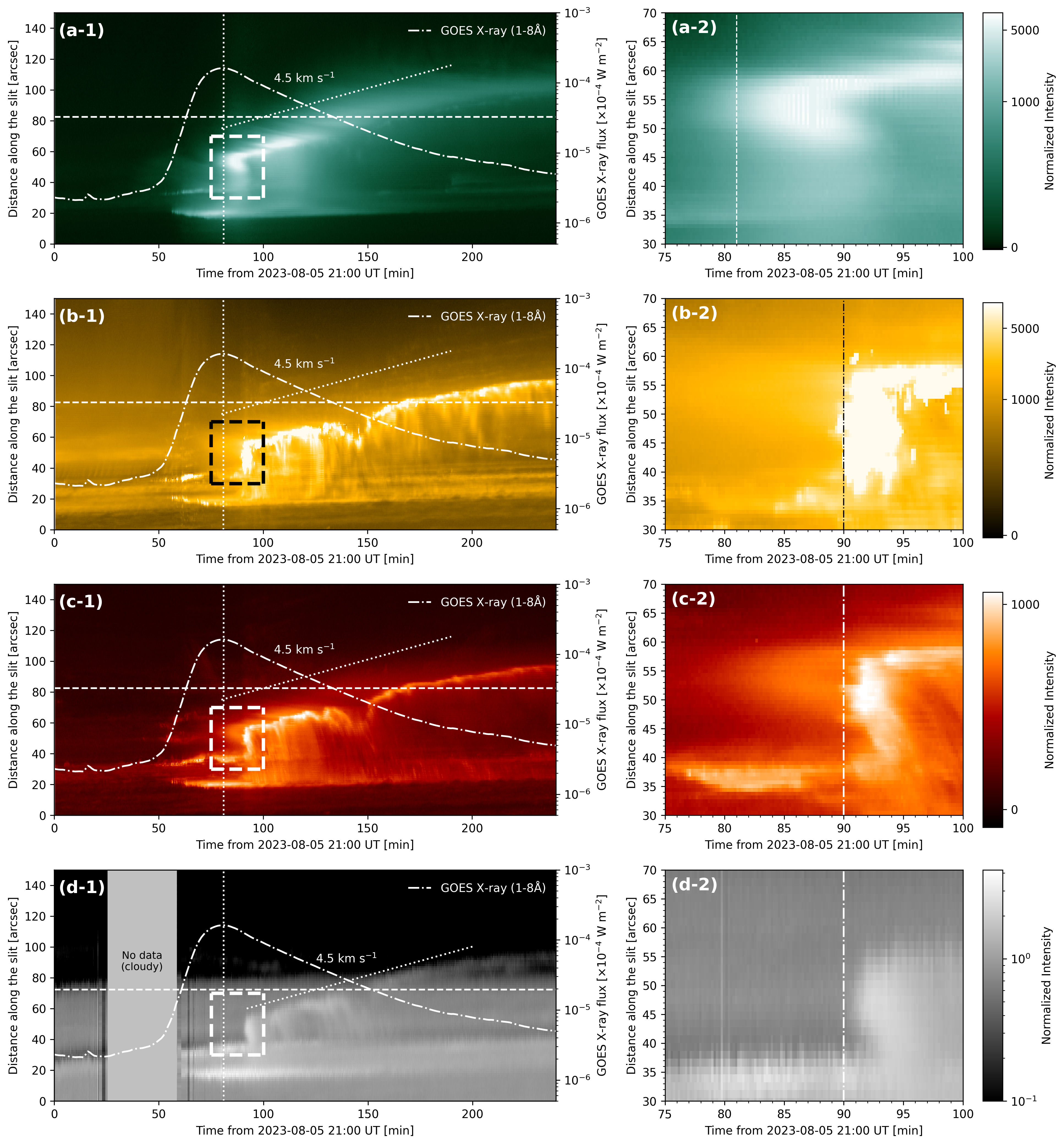}
\caption{The time-slice diagram. From panel (a-1) to (d-1), time-slice diagrams along the slit in Figure \ref{dev} ($t=210$ minutes) are shown for AIA 94 {\AA}, 171 {\AA}, 304 {\AA}, and H$\alpha$ line center, respectively. The GOES light curve is over-plotted as the white dash-dot line. The inclined white dotted lines indicate the velocity of 4.5 km s$^{-1}$. The horizontal white dashed lines indicate the solar limb, which are set as same for AIA three channels but different for the H$\alpha$ line center. Paneles (a-2)-(d-2) show the zoomed-in diagram corresponding to the white or black dashed regions in panels (a-1)-(d-1). The GOES peak time ($t=81$ minutes) is indicated by the vertical white dotted line in panels (a-1)-(d-1) and (a-2). The vertical white or black dash-dot lines in panels (b-2)-(d-2) indicate the time of $t=90$ minutes.}
\label{TH}
\end{figure}

We introduce the method of the Sun-as-a-star analysis for H$\alpha$ data \citep[see][for details]{OtsuETAL2022}.
First, we integrated H$\alpha$ spectra over the integral region $A$ including the target phenomena.
The normalizations by continuum and quiet region data were performed to suppress fluctuations in the instrument and Earth's atmospheric variations. 
Second, we subtracted pre-event data from each integrated spectrum to extract the change in spectra due to the flare.
Finally, we normalized the pre-event subtracted spectra by the full-disk integrated continuum.
The resulting normalized pre-event subtracted H$\alpha$ spectra $\Delta \tilde{S}_{H\alpha}(t,\lambda; A)$, where $t$ is time, $\lambda$ is the wavelength, and $A$ is the integral region, express the ratio of spectral changes coming from the target phenomena to the full-disk brightness of the Sun. 
In this study, the integral region for the Sun-as-a-star analysis was set as $A=A_1+A_2+A_3$ in Figure \ref{full} (c), excluding the limb ($A_{limb}$) region.
For the SDDI observation, the shift of the solar image disrupts H$\alpha$ spectra at the solar limb. This effect is purely due to the imaging and would not occur in the case of stellar observation. Thus, we excluded the limb region. We will also show the results with the limb region ($\Delta \tilde{S}_{H\alpha}(t,\lambda; A_1+A_2+A_3+A_{limb})$) just for the comparison (Figure \ref{Total_Multi} (b-1)-(b-2)). Hereafter, we call the $\Delta S_{H\alpha}(t,\lambda)=\Delta \tilde{S}_{H\alpha}(t,\lambda; A_1+A_2+A_3)$ the Sun-as-a-star H$\alpha$ spectra.
In addition to the total region $A=A_1+A_2+A_3$, we applied the above method to three regions; Region 1 ($A_{1}$), Region 2 ($A_{2}$), and Region 3 ($A_{3}$) (Figure \ref{full} (c)), and obtained $\Delta \tilde{S}_{H\alpha}(t,\lambda; A_1)$, $\Delta \tilde{S}_{H\alpha}(t,\lambda; A_2)$, and $\Delta \tilde{S}_{H\alpha}(t,\lambda; A_3)$ to investigate the contributions from flare ribbons, post-flare loops on disk, and off-limb post-flare loops (Figure \ref{ribbon_loops}).
$\Delta \tilde{S}_{H\alpha}(t,\lambda; A_1)$ mainly includes the contributions from the east ribbon ($A_1$: black region in Figure \ref{full}).
$\Delta \tilde{S}_{H\alpha}(t,\lambda; A_2)$ mainly includes the contributions from the post-flare loops, although it is also affected by the west ribbon ($A_2$: red region in Figure \ref{full}).
$\Delta \tilde{S}_{H\alpha}(t,\lambda; A_3)$ includes the contributions from the off-limb region ($A_3$: orange region in Figure \ref{full}).
We also calculated the differenced H$\alpha$ equivalent width to obtain the light curve of the H$\alpha$ line: $\Delta EW(t; A)=\int^{\mathrm{H\alpha}+\Delta\lambda}_{\mathrm{H\alpha}-\Delta\lambda}\Delta \tilde{S}_{H\alpha}(t,\lambda; A)d\lambda$ where $\Delta\lambda$ was set as $3.0$ {\AA} to include the whole spectral variations.
\section{Results} \label{sec:results}
\label{subsec:ReSDDI}

\subsection{Result of Sun-as-a-star analysis}
\begin{figure}[htbp]
\centering
\includegraphics[width=10cm]
{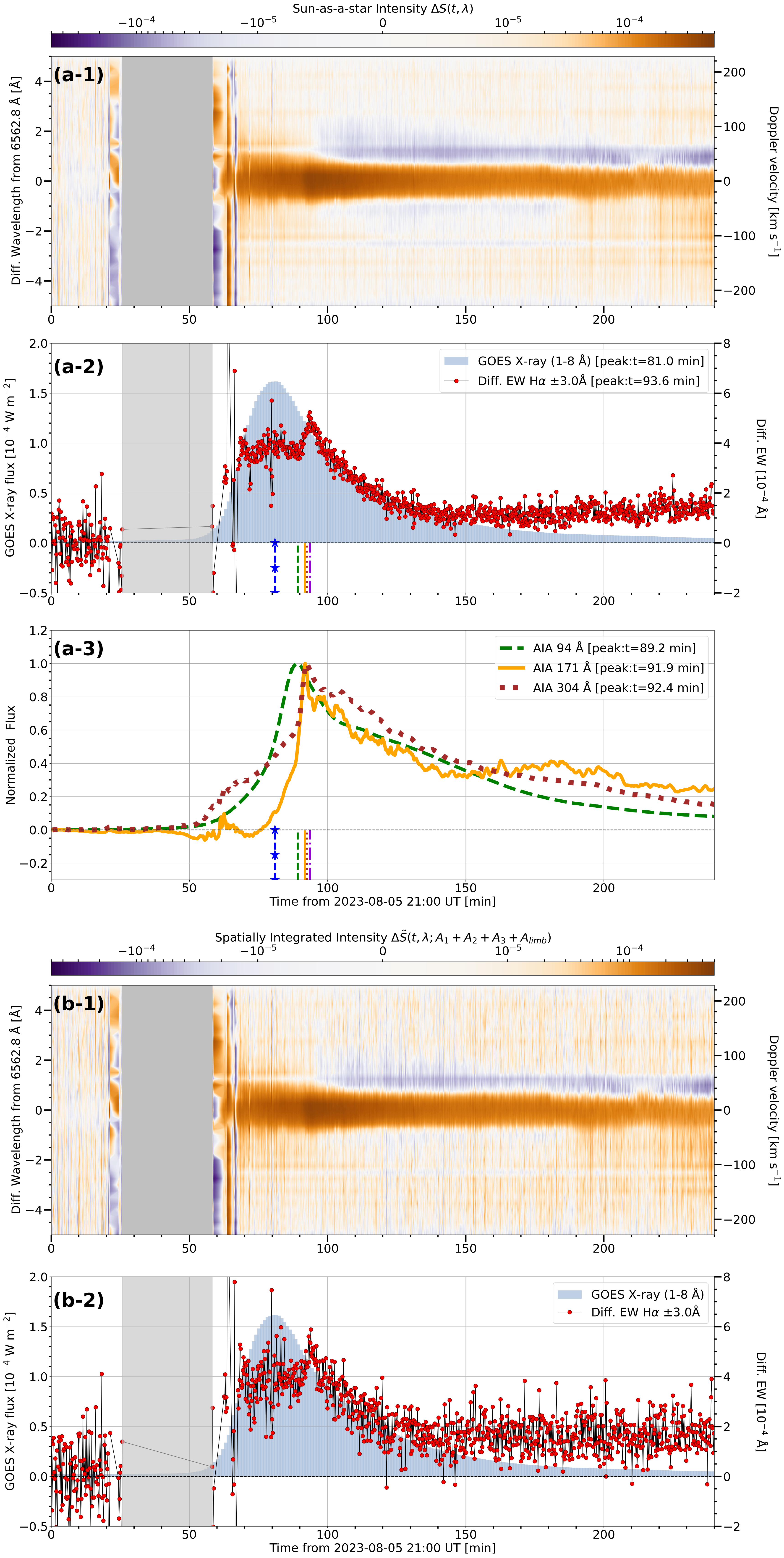}
\caption{(a-1)-(a-3) The results of the Sun-as-a-star analysis. (b-1) and (b-2) Integrated results with limb for the comparison. (a-1) The spatially integrated pre-event-subtracted H$\alpha$ spectrum normalized by the full-disk integrated continuum is shown as a two-dimensional color map for the Sun-as-a-star spectra with the total region ($A_{1}+A_{2}+A_{3}$).  Orange and purple indicate emission and absorption compared with the pre-event state, respectively. (a-2) The differenced H$\alpha$ equivalent widths and the GOES light curve are plotted as red circles and skyblue histogram, respectively. (a-3) The light curves of AIA 94, 171 ,and 304 {\AA} are shown as the green dashed, orange solid, and dark-red dotted lines, respectively. In panel (a-2) and (a-3), the vertical blue starred, green dashed, orange solid, dark-red dotted, and purple dash-dot lines indicate the peak time of the light curves of GOES, AIA 94, 171, 304 {\AA}, and H$\alpha$. The peak times are also indicated in the legends of panels (a-2) and (a-3). (b-1) and (b-2) The same as panels (a-1) and (a-2) but for results obtained over the integration involving the limb region ($A_{limb}$) in addition to the total region. The gray regions in panels (a-1), (a-2), (b-1), and (b-2) means that no data is available for the SDDI due to the bad weather condition.}
\label{Total_Multi}
\end{figure}

First, we show the result obtained by integration over the total region (Region 1+2+3) i.e. the Sun-as-a-star H$\alpha$ spectra along with the result of AIA EUV channels.
Figure \ref{Total_Multi} (a-1) shows the two dimensional color map of the Sun-as-a-star H$\alpha$ spectra $\Delta S_{H\alpha}(t,\lambda)$ obtained by the integration over the Region 1+2+3 ($A_1+A_2+A_3$).
Figure \ref{Total_Multi} (a-2) shows the light curves of the SDDI H$\alpha$ ($\Delta EW(t; A_1+A_2+A_3)$) and GOES in linear scale. 
In Figure \ref{Total_Multi} (a-3), the light curves of AIA 94, 171 and 304 {\AA} are shown. The peak time of the light curves are indicated by vertical lines in Figure \ref{Total_Multi} (a-2) and (a-3).
Figure \ref{Total_Multi} (b-1) and (b-2) are the same as (a-1) and (a-2) but for the results with the limb region.
The general trends such as time evolution of the light curves are same for the two cases, although the results with limb region become much more noisy compared with those without the limb region.
These justify that we select the results without the limb region (Figure \ref{Total_Multi} (a-1) and (a-2)) as the Sun-as-a-star results.

The Sun-as-a-star H$\alpha$ spectra and the multi-wavelength light cures show the following features:
\begin{itemize}
    \item [(i)] The two-step increase of $\Delta EW(t; A_1+A_2+A_3)$ can be confirmed. The initial increase is similar to GOES soft X-rays flux, whereas the second peak is delayed in about 13 minutes from the GOES peak.
    AIA light curves also showed delayed peaks compared with the GOES peak.
    The cooler channels and H$\alpha$ line show delayed peak times compared with the hotter ones.
    \item [(ii)] The redshifted/blueshifted absorptions can be confirmed in the Sun-as-a-star H$\alpha$ spectra, which come from the downflows along the post-flare loops. 
    \item [(iii)] The H$\alpha$ $\Delta EW(t; A_1+A_2+A_3)$ does not monotonically decrease as in GOES but stops at around $t=150$ minutes.
\end{itemize}

\subsection{Results for sub-regions}
Here, we show the results obtained by spatial integration of the H$\alpha$ spectra over Region 1-3 to investigate contributions from the flare ribbons and the on-disk/off-limb post-flare loops to the Sun-as-a-star H$\alpha$ spectra.

\begin{figure}[htbp]
\centering
\includegraphics[width=19.5cm]
{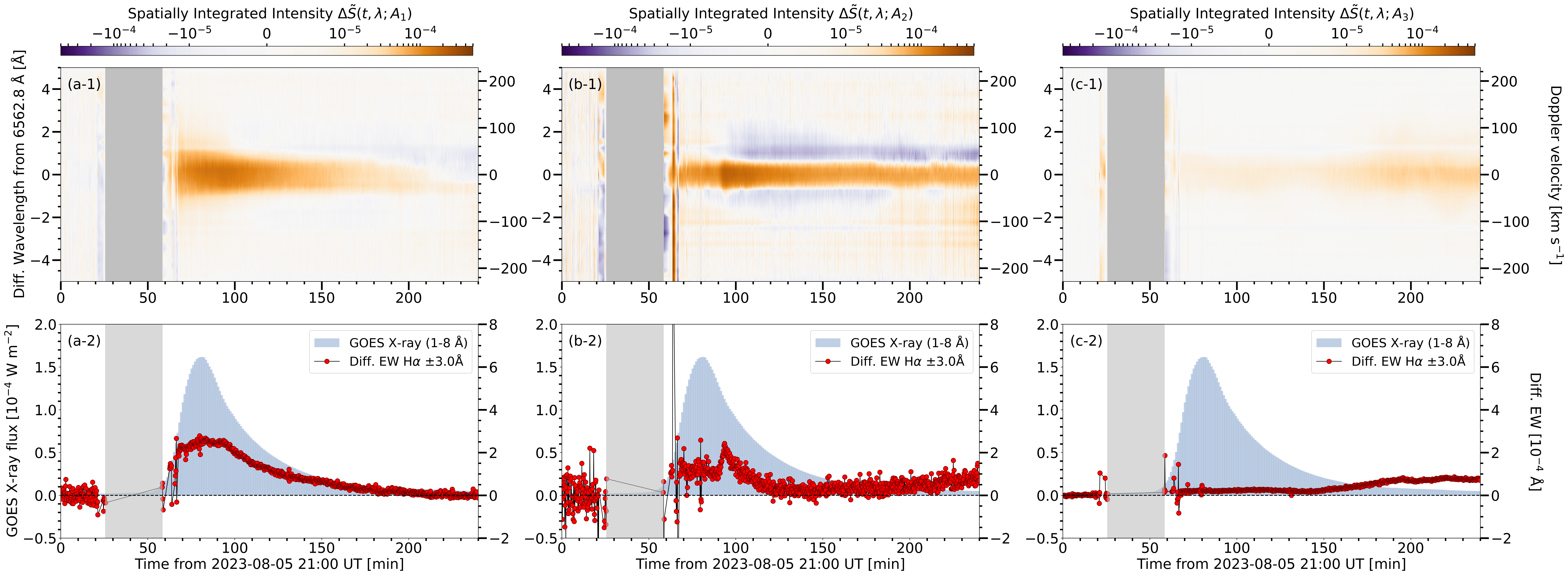}
\caption{H$\alpha$ dynamic spectra and their light curves. Panels (a-1), (b-1), and (c-1) are the same as Figure \ref{Total_Multi} (a-1) but for the Region 1, 2, and 3 (Figure \ref{full}), respectively.
In panels (a-2), (b-2), and (c-2), the differenced H$\alpha$ equivalent widths are shown for the Region 1, 2, and 3, respectively. The GOES light curve is also plotted as skyblue histogram in the panels. In the gray regions in all panels, no data is available for the SDDI due to the bad weather condition.}
\label{ribbon_loops}
\end{figure}

(\textit{Region 1})
Figure \ref{ribbon_loops} (a-1) shows the two dimensional color map of the pre-event subtracted H$\alpha$ spectra $\Delta \tilde{S}_{H\alpha}(t,\lambda; A_1)$ obtained by the integration over the Region 1. Figure \ref{ribbon_loops} (a-2) shows the light curves of $\Delta EW(t; A_1)$ of H$\alpha$ and GOES soft X-rays.
$\Delta \tilde{S}_{H\alpha}(t,\lambda; A_1)$ show the brightening near the line center corresponding to the east ribbon.
The H$\alpha$ and GOES soft X-rays light curves show the similar peak time and decay.

(\textit{Region 2})
Figure \ref{ribbon_loops} (b-1) and (b-2) are the same as Figure \ref{ribbon_loops} (a-1) and (a-2) but for Region 2.
The H$\alpha$ light curve $\Delta EW(t; A_2)$ show the two-step increase.
The first increase around $t=70-90$ minutes corresponds to the west ribbon, whereas the second increase around $t=90$ minutes comes from the post-flare loops.
Corresponding to the two-step increase of the light curve, the Sun-as-a-star H$\alpha$ spectra shows brightening near the line center.
Additionally, $\Delta \tilde{S}_{H\alpha}(t,\lambda; A_2)$ shows redshifted/blueshifted absorptions coming from the downflows along the post-flare loops.
The redshifted and blueshifted velocities are up to $100$ km s$^{-1}$ and $-80$ km s$^{-1}$, respectively. We note that these are Doppler velocities. Due to projection, true downflow velocities will be higher than these Doppler velocities.

(\textit{Region 3})
Figure \ref{ribbon_loops} (c-1) and (c-2) are the same as Figure \ref{ribbon_loops} (a-1) and (a-2) but for Region 3.
The H$\alpha$ light curve $\Delta EW(t; A_3)$ show the increase around $t=150$ minutes, which comes from the off-limb post-flare loops.
Correspondingly, $\Delta \tilde{S}_{H\alpha}(t,\lambda; A_3)$ also shows the brightening near the line center.

\section{Discussion and Conclusions} \label{sec:discussion}
\subsection{Dynamics of Post-flare Loops in the Sun-as-a-star Data}
\subsubsection{Cooling: Peak Time Difference in Multi-temperature Light Curves}
As shown in Figure \ref{ribbon_loops} (a-1) and (a-2), the east ribbon inside the Region 1 provides the enhancement of the H$\alpha$ light curve which has a peak and decay similar to the GOES flux.
Corresponding to the appearance of the H$\alpha$ post-flare loops, the H$\alpha$ light curve for Region 2 exhibited the delayed peak ($t=93.6$ minutes) compared with the GOES peak ($t=81$ minutes).
However, the west ribbon provides the initial increase even in the case of Region 2 (Figure \ref{ribbon_loops} (b-1)-(b-2)).
Thus, the initial and secondary increase in the Sun-as-a-star result comes from the flare ribbons and post-flare loops, respectively (Figure \ref{Total_Multi} (a-1) and (a-2)). 
Our result showed that post-flare loops can lead to non-negligible enhancement in H$\alpha$ compared to flare ribbons.
The EUV light curves also showed the delayed peak compared to the GOES flux.
The peaks appear in the order of GOES soft X-ray ($t=81$ minutes), AIA 94 ($t=89.2$ minutes), 171 ($t=91.9$ minutes), 304 ($t=92.4$ minutes), and H$\alpha$ ($t=93.6$ minutes), which means the cooling of the post-flare loops can be confirmed as peak time difference even in the spatially integrated data.
The peak time differences from GOES, AIA 94, 171, and 304 to H$\alpha$ are approximately 13, 4.4, 1.7, and 1.2 minutes, respectively (Figure \ref{Total_Multi} (a-3))

In this flare, the downflows of cooled and condensed plasma were observed (Figure \ref{dev}, see also Section \ref{downflows}), which implies the radiative loss is effective in the cooling process (initially the conductive cooling may dominate in the hot loops).
To investigate whether the delayed peaks represent the radiative cooling or not, we compared the peak time differences with the radiative cooling time scale.
The radiative cooling time scale is calculated using the following equation:
\begin{equation}
 \tau_{rad}=3n_e kT/(n_e ^2Q(T)),
\end{equation}
where $n_e$, $T$, $k$, and $Q(T)$ [erg cm$^3$ s$^{-1}$] are electron density, electron temperature, Boltzmann constant, and radiative cooling function, respectively.
We used the typical electron density $n_e=10^{11}$ cm$^{-3}$ for post-flare loops \citep{Kamio2003SoPh} and calculated radiative cooling function $Q(T)$ using CHIANTI 10.1 \citep{Dere1997A&AS..125..149D, Dere2023ApJS..268...52D} with coronal abundance.
The cooling times for $T=10^7$K (GOES), $10^{6.8}$ K (94 {\AA}), $10^{5.9}$ K (171 {\AA}), and $10^{4.9}$ K (304 {\AA}) are calculated to be $\tau_{rad}\sim14, 7, 0.2, 0.01$ minutes, respectively\footnote{With photospheric abundance, the cooling times for $T=10^7$K (GOES), $10^{6.8}$ K (94 {\AA}), $10^{5.9}$ K (171 {\AA}), and $10^{4.9}$ K (304 {\AA}) are calculated to be $\tau_{rad}\sim31, 17, 0.5, 0.02$ minutes, respectively.}. We confirmed that the electron density of the post-flare loops is approximately $10^{10.7}$ cm$^{-3}$ using AIA/DEM analysis \citep{Hannah2012A&A...539A.146H}.
We also calculated cooling times using the method in \citet{Svestka1987SoPh..108..411S} and \citet{Schmieder1995SoPh..156..337S}, which provided the cooling times comparable to the estimated $\tau_{rad}$.
The peak time differences from GOES and AIA 94 to H$\alpha$ ($\sim13, 4.4$ minutes) are close to $\tau_{rad}\sim14, 7$ minutes, whereas those for AIA 171 and 304 ($\sim1.7, 1.2$ minutes) are much larger than $\tau_{rad}\sim0.2, 0.01$ minutes.
During the evolution of post-flare loops, electron density could change.
This may be the cause of the inconsistency in AIA 171 and 304.
To estimate the radiative cooling time scale and compare it with peak time difference more accurately, measuring the time development of electron density in post-flare loops is critical.
In this rough estimation, the coronal abundance is used for $Q(T)$ but abundance should be carefully treated because the plasma fills the post-flare loops through chromospheric evaporation.
Moreover, plasma of $10^6$ K affect the formation of AIA 304 \citep{ODwyer2010A&A...521A..21O}, although the typical response temperature of AIA 304 is $\sim10^{4.9}$ K.
This may also make the peak time difference from 304 to H$\alpha$ larger than the calculated $\tau_{rad}$.
We note that conductive cooling may also be effective for cooling of hot loops and make the cooling time shorter than our estimation. Additionally, we should consider time variation of temperature for more accurate estimate of cooling time. We will consider these factors in further investigations using multiple flare events.

\subsubsection{Downflows: Redshifted and Blueshifted Absorption in H$\alpha$ Spectra}\label{downflows}
The H$\alpha$ dynamic spectrum for Region 2 shows the redshifted/blueshifted absorption from the downflows along the post-flare loops, which can be confirmed even in the Sun-as-a-star H$\alpha$ spectra (Figure \ref{ribbon_loops} (b-1)).
As described in Section \ref{sec:obs}, the post-flare loops are located near the solar limb and tilted against the line-of-sight direction.
As a result, the downflows are observed as a blueshifted component and a redshifted one in the H$\alpha$ imaging observation, and these redshifted/blueshifted absorptions appeared even in the Sun-as-a-star spectra.
Unlike the present flare, the previous Sun-as-a-star study reported only the redshifted absorption --which is probably related to downflows along post-flare loops-- during the decay phase of an M1.1 flare which occurred relatively close to the disk center \citep[Event 4 in][]{OtsuETAL2022}.
The Sun-as-a-star results of the present X1.6 and previous M1.1 flares suggest that a flare with blueshifted absorptions from post-flare loops is likely to occur near a solar/stellar limb.
The dependence of shifted absorptions caused by downflows on occurrence locations should be investigated with a model of flows inside loops \citep[e.g.,][]{Ikuta2024ApJ...963...50I}.
We note that dependence of red and blue asymmetry related to post-flare loops on flare locations are also discussed in \citet{Wollmann2023A&A...669A.118W} for observations on an M dwarf star.
Furthermore, it will be important to compare locations of stellar post-flare loops based on signatures of downflows and those of starspots deduced from starspot mapping \citep[e.g.,][]{Ikuta2020ApJ...902...73I, Ikuta2023ApJ...948...64I} for ensuring scenarios of stellar flares.

\subsubsection{Rise of the Loops: Stop of the H$\alpha$ Decay}
The H$\alpha$ light curve for Region 3 begins to increase around $t=150$ minutes, corresponding to the appearance of the off-limb post-flare loops which is a consequence of the generation of higher post-flare loops (Figure \ref{ribbon_loops} (c-2)). 
Off-limb loops have no background intensity and they are fully in emission. Thus, they would exhibit stronger enhancement than on-disk loops compared to the pre-event state.
As a result, the Sun-as-a-star H$\alpha$ light curve showed the stop of the decay around $t=150$ minutes.
Our result showed that the different contrast to the background between the cases of on-disk and off-limb loops can lead to tracing the plane-of-sky motions of plasma even in spatially integrated data. 
Additionally, the appearance of off-limb loops reflects that the flare occurred near the solar limb.
Therefore, the stop of the decay of H$\alpha$ light curves can also be useful to deduce the locations of stellar flares. 

\subsection{Conclusion and Implications for Stellar Observations}
In this study, we performed the Sun-as-a-star analysis of the X1.6 flare on 2023 August 5 which exhibited the post-flare loops in EUV channels and H$\alpha$ line spectra. 
We found that even the Sun-as-a-star data showed three characteristics of post-flare loops. 
\begin{itemize}
    \item [(1)] The H$\alpha$ light curve showed two distinct peaks corresponding to the flare ribbons and the post-flare loops. The plasma cooling in the post-flare loops generated different peak times in soft X-rays, EUV, and H$\alpha$ light curves.
    \item [(2)] Downflows were confirmed as simultaneous redshifted/blueshifted absorptions in Sun-as-a-star H$\alpha$ spectra. 
    \item [(3)] The apparent rise of post-flare loops was recognized as a slowing of the decay for the H$\alpha$ light curve.
\end{itemize}
Our results are crucial to investigate stellar post-flare loops in spectroscopic and multi-wavelength observations.
Additionally, we emphasize that signatures of post-flare loops would reflect the occurrence location of flares. Statistical studies on solar post-flare loops are crucial for further understanding of their dependence on flare locations.

\begin{acknowledgments}
The authors thank the anonymous referee for constructive comments that significantly improved the quality of this Letter.
We express our sincere gratitude to the staff of Hida Observatory for developing and maintaining the instrument and daily observation. 
We would like to acknowledge the data use from GOES and SDO.
SDO is a mission for NASA’s Living With a Star program. 
This work was supported by JSPS KAKENHI grant Nos. JP24K07093  (PI: A. A.), JP21H01131 (PI: K. S.), JP24K00680 (PI: K. S.), JP24K17082 (PI: K. I.), and JP24H00248 (PI: D. Nogami).
This work was also supported by JST SPRING, Grant Number JPMJSP2110 (T.O.). 
\end{acknowledgments}

\software{sunpy \citep{Sunpy2020ApJ}, ChiantiPy \citep{Dere2013ascl.soft08017D}}

\clearpage


\end{document}